\documentclass[conference]{IEEEtran}
\IEEEoverridecommandlockouts
\usepackage{graphicx}
\usepackage{listings}
\usepackage{fancyvrb}
\usepackage[hyphens]{url}
\usepackage{color}
\usepackage{cite}
\usepackage{flushend}

\lstnewenvironment{mylisting}[1][]
    {\lstset{
        basicstyle={\ttfamily},
        identifierstyle={\small},
        commentstyle={\smallitshape},
        keywordstyle={\small\bfseries},
        ndkeywordstyle={\small},
        stringstyle={\footnotesize\ttfamily},
        frame={tb},
        breaklines=false,
        columns=[l]{fullflexible},
        numbers=none,
        numberstyle={\scriptsize},
        stepnumber=1,
        lineskip=-0.5ex
    }
}{}

\begin{document}
\title{GbHammer: Malicious Inter-process Page Sharing by Hammering Global Bits in Page Table Entries}

\author{
    \IEEEauthorblockN{
    Keigo Yoshioka\IEEEauthorrefmark{1},
    Soramichi Akiyama\IEEEauthorrefmark{2}}
    \IEEEauthorblockA{\IEEEauthorrefmark{1}Shibuya Junior \& Senior High School, Tokyo, Japan}
    \IEEEauthorblockA{\IEEEauthorrefmark{2}College of Information Science and Engineering, Ritsumeikan University, Osaka, Japan}
    \IEEEauthorblockA{Email: s-akym@fc.ritsumei.ac.jp}
}

\maketitle              
\begin{abstract}
RowHammer is a vulnerability inside DRAM chips where an attacker repeatedly accesses a DRAM row to flip bits in the nearby rows without directly accessing them.
Several studies have found that flipping bits in the address part inside a page table entry (PTE) leads to serious security risks such as privilege escalation.
However, the risk of management bits in a PTE being flipped by RowHammer has not yet been discussed as far as we know.
In this paper, we point out a new vulnerability called {\it GbHammer} that allows an attacker to maliciously share a physical memory page with a victim by hammering the global bit in a PTE.
GbHammer not only creates a shared page but also enables the attacker to (1) make the victim's process execute arbitrary binary and (2) snoop on the victim's secret data through the shared page.
We demonstrate the two exploits on a real Linux kernel running on a cycle-accurate CPU simulator.
We also discuss possible mitigation measures for GbHammer and the risk of GbHammer in non-x86 ISAs.
\end{abstract}


\section{Introduction}
RowHammer is a vulnerability inside DRAM chips where an attacker repeatedly accesses a DRAM row to flip bits in the nearby rows without directly accessing them~\cite{Kim2014}.
It can result in serious outcomes such as OS-level privilege escalation~\cite{Zhang2020}, secret data leakage~\cite{Kwong2020},
 Denial-of-Service~\cite{Jang2017}, breaking VMM-level memory isolation~\cite{Xiao2016}, and confusing AI models~\cite{Li2024}.

Despite many RowHammer-based exploits found,
the risk of management bits in page table entries (PTEs) being hammered is yet to be discussed.
To this end, we demonstrate two new exploits when the global bit of a PTE is flipped by RowHammer, which we refer to as {\it GbHammer}.
GbHammer forces the CPU to use the same address translation information between the attacker and the victim,
resulting in a physical memory page maliciously shared by them.
Through the shared page, the attacker can make the victim execute binary code that the attacker crafts and snoop on secret data of the victim.

The contributions of this paper are as follows:
\begin{enumerate}
\item We are the first to discuss the risk of management bits in PTEs being hammered as far as we know.
\item We propose two new exploits based on GbHammer, and demonstrate that they indeed work using a cycle-accurate CPU simulator gem5 and a real Linux kernel.
\item We investigate the specifications of ISAs other than x86\_64, namely ARMv7 and RISC-V, and discuss that the same exploits can be executed on them.
\item We discuss possible mitigation measures of GbHammer and the challenges in achieving these measures.
\end{enumerate}

\section{Preliminaries}
A page table entry (PTE) refers to one line of a page table that consists of a mapping from a virtual address to a physical address and some additional management bits.
Among these management bits, a PTE in x86 has one called a {\it global bit}~\cite{IntelManual}.
A PTE with the global bit enabled is called a {\it global PTE}.
A global PTE indicates that the address translation information represented by that PTE may be used among different processes.
It is beneficial for some special cases where multiple processes use the same address translation information.
For instance, the kernel address space is often mapped to the same virtual address ranges of different processes.

On Intel processors, a global PTE is associated with a {\it global TLB entry}.
TLB (Translation Look-aside Buffer) caches address translation information that is recently used and the cached information is referred to as TLB entries.
It greatly improves CPU performance because accessing the TLB is faster compared to a normal page table walk that requires multiple DRAM accesses.
When the address translation information within a global PTE is cached to the TLB, the TLB entry also becomes global.
This means that the CPU may use the same TLB entry for different processes from the one that has created the entry.
More concretely, the Intel manual~\cite{IntelManual} says that a
{\it processor may use a global TLB entry to translate a linear address, even if the TLB entry is associated with a PCID different from the current PCID}.
Here, PCID (Process-Context Identifier) is a value assigned to each TLB entry and it is unique to the process that created the entry.
PCID prevents the TLB from being completely flushed in a context switch.

\section{New Vulnerability: GbHammer}
\subsection{Threat Model}
In this paper, we assume that an attacker can login to the same physical machine as the victim and can execute programs with a user privilege.
Specifically, an attacker can
\begin{enumerate}
    \item login to the same machine as the victim by using measures such as an SSH client,
    \item prepare a program on that machine by either writing code and compiling it or transferring an already compiled binary by using measures such as SCP,
    \item execute the prepared program with a user privilege on any core including the one on which the victim’s process is executed, and
    \item access the executable file that the victim invokes as a process (the intention is explained in Section~\ref{section:steps_of_gbhammer}).
\end{enumerate}

\subsection{\label{section:Detail of the Attack}Malicious Page Sharing} 
\begin{figure}[t]
\begin{center}
\centering
\includegraphics[width=\columnwidth]{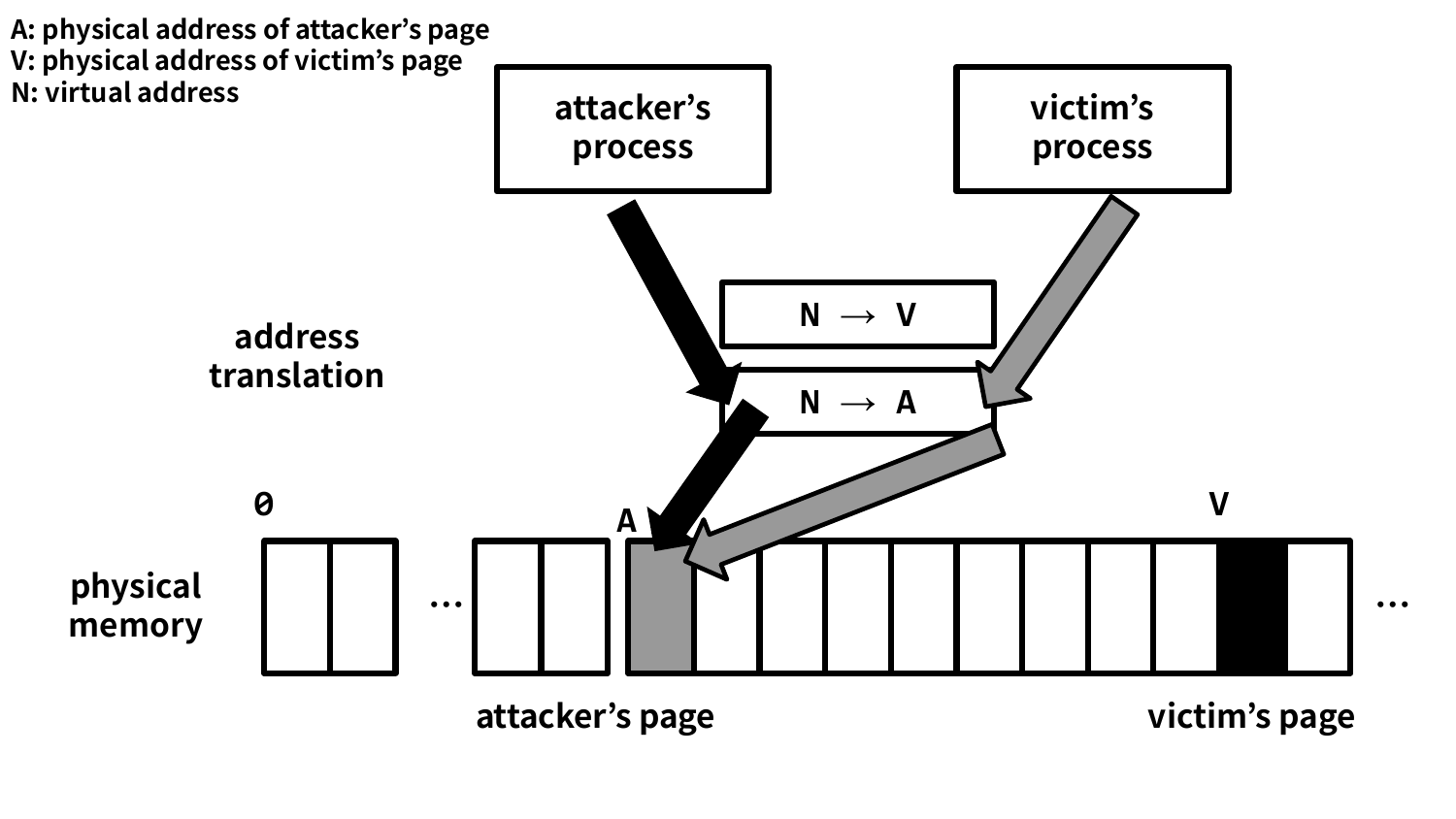} 
\caption{\label{fig:tlb_compromised}Malicious Page Sharing by GbHammer}
\end{center}
\end{figure}

GbHammer is an attack that allows an attacker's process to maliciously share physical memory pages with a victim's process. This attack is achieved by flipping a global bit of a PTE by RowHammer.
This makes the same address translation information used by both the attacker's and the victim's processes, resulting in a shared physical page among them.

Fig.~\ref{fig:tlb_compromised} illustrates the situation where a physical page is maliciously shared by GbHammer.
Here, \verb|N| represents a virtual address that both the attacker's and victim's processes use.
Note that using the same virtual address itself is completely normal and allows no malicious data access thanks to the isolation of virtual address spaces.
\verb|A| and \verb|V| represent the physical addresses that are mapped from \verb|N| in the virtual address spaces of the attacker's process and the victim's process, respectively.
These mappings are shown in the figure as the {\it address translation} labeled with \verb|N|$\rightarrow$\verb|A| and \verb|N|$\rightarrow$\verb|V|, respectively.
The black arrows from the attacker's process to the physical page at address \verb|A| show that memory accesses by the attacker's process go to the attacker's page as expected.
On the other hand, the grey arrows from the victim's process show that memory accesses from the victim's process are maliciously forwarded to the {\bf attacker's} page.

In summary, GbHammer allows an attacker to create a shared page as follows:
\begin{enumerate}
    \item The attacker's process flips the global bit of the PTE that maps \verb|N| to \verb|A| in their own address space using RowHammer. This makes this PTE global.
    \item The attacker's process issues a memory access to virtual address \verb|N|. This creates a global TLB entry that maps \verb|N| to \verb|A|.
    \item The attacker waits until the victim's process accesses virtual address \verb|N|. Because the CPU has a global TLB entry that maps \verb|N| to \verb|A|, the access by the victim's process ends up going to the physical address \verb|A| instead of \verb|V|.
    \item This means that the physical page starting from physical address \verb|A| is now shared among the attacker's process and the victim's process.
\end{enumerate}


\subsection{\label{section:steps_of_gbhammer}Steps of GbHammer}
Here we elaborate on the steps of GbHammer in more detail.
\begin{itemize}
    \item {\bf Step (1)}: The attacker acquires a virtual address \verb|N| that the victim's process uses.
    This is necessary because GbHammer only works when the attacker's and the victim's processes use the same virtual address.
    \item {\bf Step (2)}: The attacker creates a mapping from \verb|N| to any physical address in the attacker's virtual address space.
    \item {\bf Step (3)}: The attacker flips the global bit in the PTE that maps \verb|N| in the attacker's virtual address space by RowHammer.
\end{itemize}

{\bf Step (1)}: The attacker’s process conducts the following two procedures.
Note that the procedures described here only work for the Arbitrary Binary Execution exploit based on GbHammer (explained in Section~\ref{section:exploits}),
and how to achieve Step~(1) for the Data Snooping exploit is future work.

First, the attacker acquires the offset of binary code that the victim's process executes.
This is done by disassembling the binary that is executed by the victim’s process.
Although the attacker only has a normal user privilege on the target machine,
they can still achieve this when the victim executes a pre-compiled binary downloaded from software repositories (e.g., by the \verb|apt| command).
This is a common assumption in some ROP-style attacks~\cite{Muntean2021,Schuster2015}.

Second, the attacker bypasses ASLR (Address Space Layout Randomization) to acquire the starting virtual address on which the code of the victim's process is placed.
By combining this address and the offset acquired in the previous paragraph, the attacker can know the virtual address \verb|N| of an arbitrary piece of code that the victim executes.
To bypass ASLR, the attacker can leverage existing techniques such as detecting collisions in the BTB (Branch Target Buffer)~\cite{Evtyushkin2016}.

{\bf Step (2)}: There are two methods to achieve Step~(2).
The first method is to use the \verb|mmap| systemcall.
POSIX.1-2001 defines the prototype of \verb|mmap| as follows~\cite{mmap_manpage}.

\begin{verbatim}
void *mmap(void *addr, size_t length,
int prot, int flags, int fd, off_t offset);
\end{verbatim}

\verb|addr| is a hint to the OS for the starting address of the created mapping.
The attacker can specify \verb|N| to this argument and the OS usually respects it when creating a mapping.

The second method to achieve Step~(2) is to specify \verb|N| as an address of functions and global variables inside an ELF binary.
An ELF binary file contains virtual addresses on which sections are loaded, and this address is used as-is when the binary is compiled with the \verb|-static| option of gcc.

{\bf Step (3)}:
The attacker's process conducts the following five procedures.
First, the attacker allocates a large and continuous memory region $R$ with \verb|mmap|.
The returned memory region is also continuous on the physical memory due to the characteristics of the buddy allocator of Linux~\cite{Kwong2020,van2016}.
A corner case scenario where memory is fragmented is also discussed in~\cite{Kwong2020}.
Second, the attacker hammers pages in $R$ to find a {\it target page}.
A target page is a page that has a bit vulnerable to RowHammer at the bit position which would be interpreted as global bits when the page is used as a page table page.
For example, when \verb|N| is \verb|0x20000|, any page whose 32712 ($= 64 \times 511 + 8$)$^{\rm th}$ bit from the least significant bit is vulnerable can be a target page.
Note that the address translation information for virtual address \verb|0x20000| is placed in the first PTE out of 512 PTEs (64 bits each) stored in a page table page.
Third, the attacker returns the target page to the OS by \verb|munmap|.
Fourth, the attacker creates a mapping from virtual address \verb|N| using \verb|mmap| with the \verb|MAP_POPULATE| flag.
Because the OS reuses the just-returned page as the page table page for the newly created mapping~\cite{project_zero_rowhammer},
this makes the target page to be used to store the PTE that maps virtual address \verb|N| to a physical page.
Finally, the attacker uses other pages in $R$ to hammer the target page until the global bit flips.
A bit that has flipped in the previous hammering is likely to flip again due to the strong locality of bit-flips.

We make two important notes here.
{\bf First}, targeting a management bit of a PTE with RowHammer is almost equivalent to targeting the address part (the only difference is the bit position to hammer),
which is already done in a body of work~\cite{project_zero_rowhammer,van2016,Xiao2016,Zhang2020}.
Therefore, we believe that targeting a global bit is feasible although our experiments focus on the outcomes of it.
{\bf Second}, separating user- and kernel-space pages in the physical memory to prevent the OS from reusing a target page as a page table page can be invalidated by PTHammer~\cite{Zhang2020}.
As far as we know, there is no solution yet to PTHammer except for activation-tracking mechanisms~\cite{Bostanci2024,Saxena2024} that require additional hardware.

\subsection{\label{section:exploits}Exploits based on GbHammer}
GbHammer enables two new exploits: arbitrary binary execution and data snooping.

\subsubsection{Arbitrary Binary Execution}
\begin{figure}[h]
\begin{center}
\centering
\includegraphics[width=1.15\columnwidth]{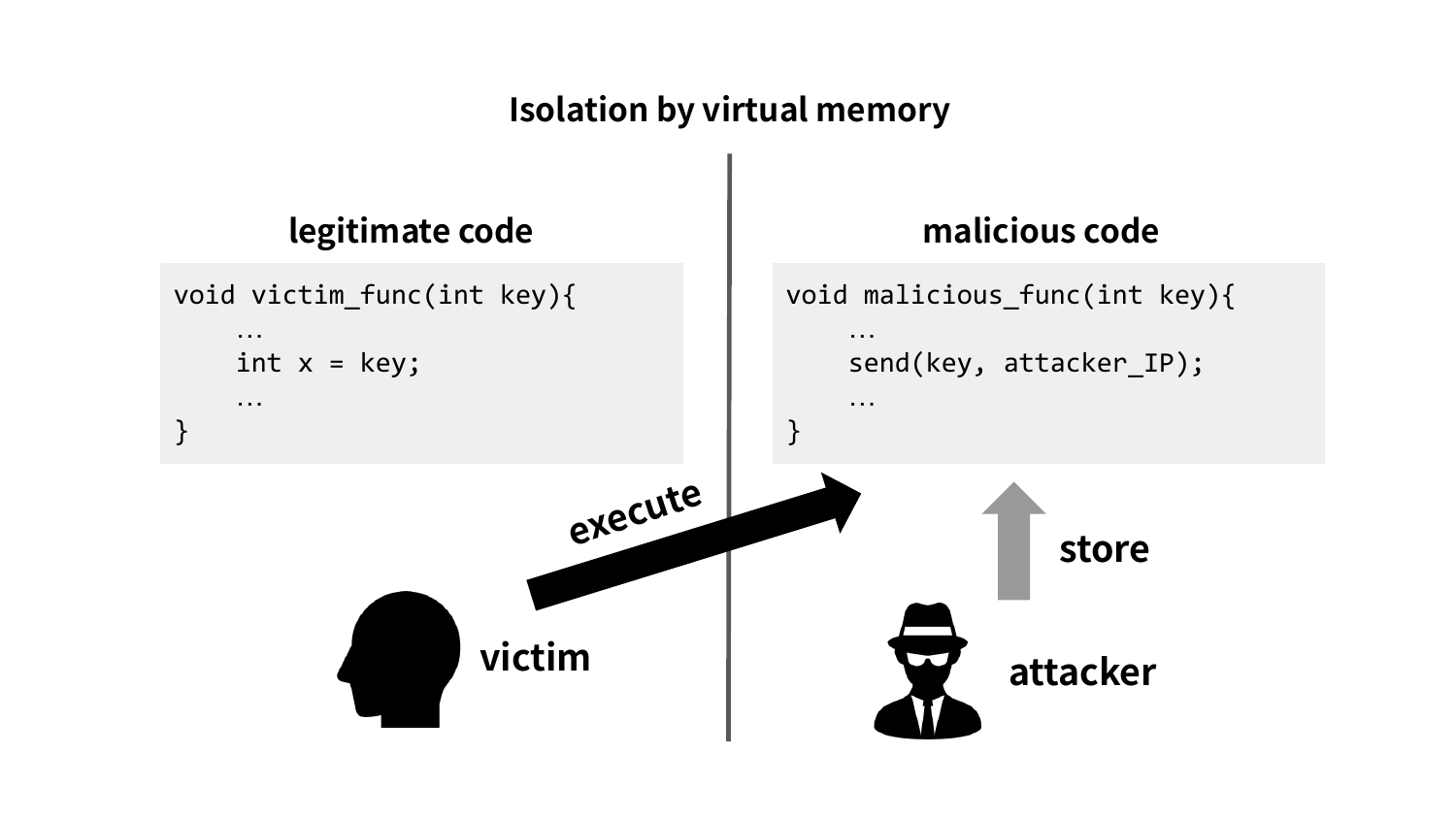} 
\caption{\label{fig:binary_exploit}Binary Execution Exploit}
\end{center}
\end{figure}

The attacker's process can make the victim's process execute an arbitrary binary as illustrated in Figure~\ref{fig:binary_exploit}.
To do this, the attacker's process maliciously creates a share page that is mapped from virtual address \verb|N| where a part of the binary code of the victim's process resides in the victim's address space.
Then, the attacker's process stores the binary that it wants the victim's process to execute to the created shared page.
Third, the attacker executes the binary by itself to create a global {\bf iTLB} entry.
When the victim's process tries to execute the code at virtual address \verb|N| later on,
it will mistakenly execute the binary stored by the attacker.

\subsubsection{Data Snooping}
The attacker's process can snoop on data written by the victim's process.
To do this, the attacker's process maliciously creates a share page that is mapped from virtual address \verb|N| to which the victim's process stores secret data.
Then, the attacker's process accesses address \verb|N| to create a global {\bf dTLB} entry.
The attacker waits until the victim's process mistakenly stores its data on the shared page by writing it to address \verb|N|.
After that, the attacker simply reads the data from the shared page.

\section{Experimental Results}
\begin{table}[t]
  \caption{Experiment environment}
  \label{table:software_version}
  \centering
  \begin{tabular}{c|c}
    \hline
    Software  & Version  \\
    \hline
    Simulated OS \hspace{1mm}& \hspace{1mm}Ubuntu 18.04.2 (Linux kernel 5.4.49) \\
    gem5 & 23.0.0.1 \\
    gcc & 9.4.0 \\
    \hline
  \end{tabular}

\vspace{5mm}

  \caption{Configuration Parameters of gem5}
  \label{table:gem5_configuration}
  \centering
  \begin{tabular}{c|c}
    \hline
    Software  & Version  \\
    \hline
    Simulated ISA & x86\_64 \\
    CPU & AtomicSimpleCPU, 1 core  \\
    Memory & Atomic \\
    TLB & 64 entries each for iTLB and dTLB\\
    \hline
  \end{tabular}
\end{table}


\subsection{Reproducing GbHammer with gem5}
We reproduce GbHammer on a cycle-accurate CPU simulator gem5~\cite{gem5} and its Full System mode that can run an entire OS to assess the risk under realistic settings.
We use the latest Linux kernel known to work on it~\cite{gem5_fs_tutorial} and other software as shown in Table~\ref{table:software_version}.
We employ the \verb|AtomicSimpleCPU| model that does not simulate the detailed timing of each instruction to achieve fast simulation (tens of minutes to boot Ubuntu).
The use of this simple CPU model should not affect the validity of the experimental results because the detailed timings such as pipeline stalls are not relevant to this work.
The simulated CPU has a single core to ensure that the attacker's and the victim's processes are executed on the same core and use the same TLB.
For the memory model, we also employ a simple one named \verb|Atomic|.
The use of this simple memory model should neither affect the validity of the experimental results because we do not reproduce RowHammer inside the simulated memory unlike Hammulator~\cite{Thomas2023} does.
Other configuration parameters of gem5 are shown in Table~\ref{table:gem5_configuration}.

Reproducing GbHammer follows the procedures below.
\begin{enumerate}
    \item We fix the target virtual address (i.e., \verb|N|) to \verb|0x20000|.
    This avoids the necessity of achieving Step~(1) and allows us to focus on the outcomes of GbHammer.
    \item Processes of the attacker and the victim are invoked on the simulated OS controlled via Telnet.
    We explain how these processes are implemented later.
    \item When a TLB entry is created for the target virtual address, the entry is forcefully set as global.
    We implement this in the source code of gem5, namely \verb|src/arch/x86/tlb.cc|.
    The \verb|insert| function is modified to set the global bit to 1 when the virtual address of a new TLB entry is \verb|0x20000|.
\end{enumerate}

The choice of using gem5 to reproduce GbHammer comes from the reliability.
Although it is possible to reproduce RowHammer (and thus GbHammer) on a real machine,
reproducing it reliably requires much engineering effort such as reverse-engineering the DRAM address mapping embedded in the memory controller using existing methods~\cite{Pessl2016,Helm2020}.
We use gem5 to reliably reproduce GbHammer to focus more on the outcome and the risk of it,
rather than on how to make it happen which is based on well-established building blocks.

\subsection{Setup: Arbitrary Binary Execution}
\begin{figure}[t]
\begin{mylisting}
int f() {
    return 1;
}

void main() {
    sleep(5); // GbHammer happens here
    printf("This is victim.\n");
    printf("Expected output: 1\n");
    printf("Actual output: %d\n", f());
}
\end{mylisting}
\caption{\label{fig:binary_code}Victim Code for the Binary Execution Experiment}
\end{figure}

In this experiment, the source code of the victim's process defines a function \verb|f| that returns 1.
The victim's process sleeps for 5 seconds, calls \verb|f|, and then prints the result returned by \verb|f|.
Fig.~\ref{fig:binary_code} shows the source code of the victim's process.
The sleep duration (5 s) can be any value as long as there is enough time for the attacker to do its job between the invocation of the victim's process and the call of \verb|f|.
The attacker's process, meanwhile, creates a shared page whose virtual address starts from \verb|0x20000| by GbHammer,
writes to the shared page a binary blob compiled from a function that returns {\bf 2}, and then executes the binary blob as a function.
This creates a {\bf global} iTLB entry (because of our modification to gem5) that translates the virtual address \verb|0x20000| into the physical address of the attacker's page that contains the binary blob.

\begin{figure}[t]
\begin{center}
\centering
\includegraphics[width=0.975\columnwidth]{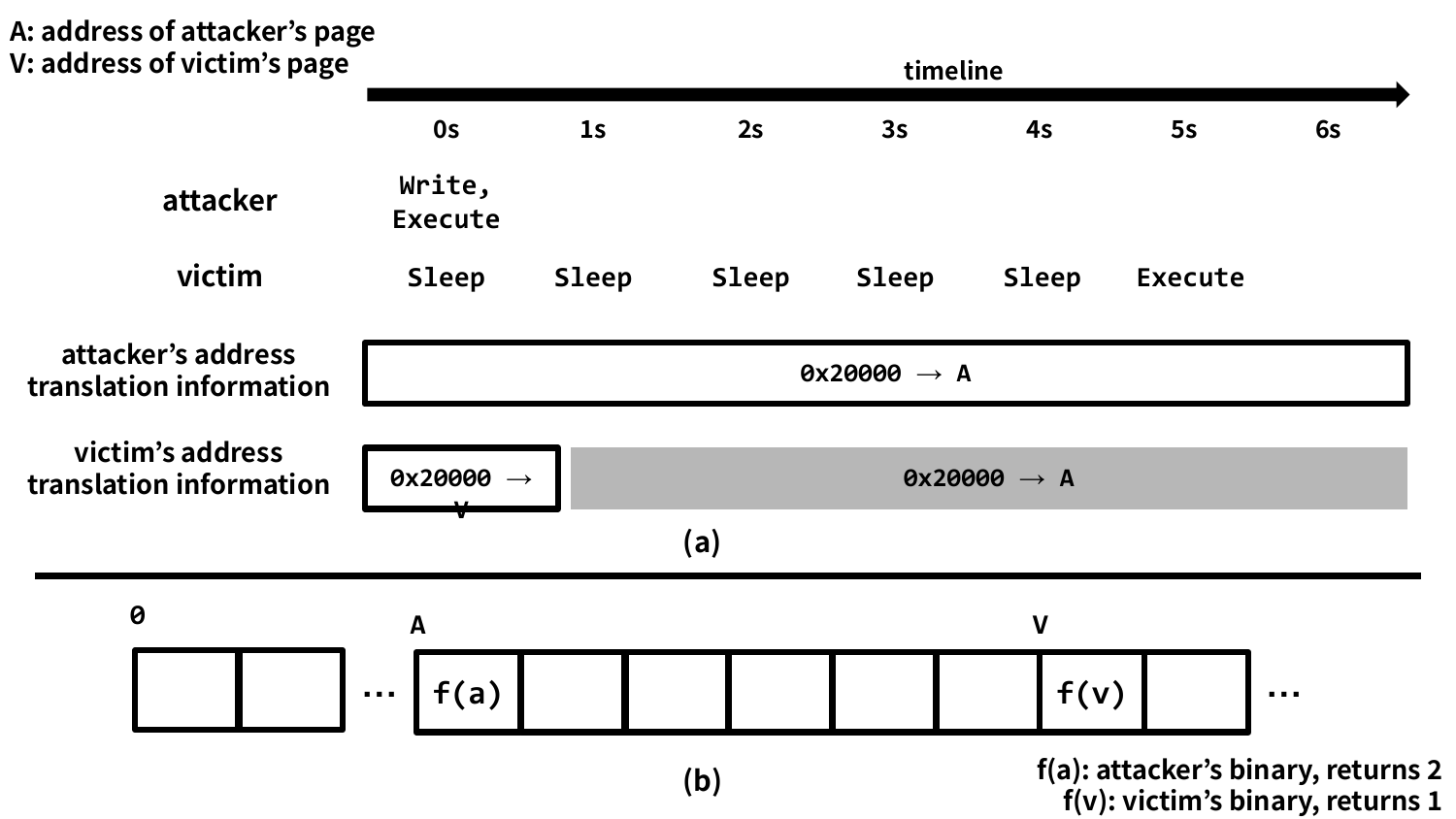} 
\caption{\label{fig:timeline_binary}Timeline of the Binary Execution Exploit}
\end{center}
\end{figure}

The victim's function \verb|f| is placed at virtual address \verb|0x20000| with the help of a linker script we craft.
A linker script is passed to \verb|gcc| and specifies how functions and global variables are placed in the virtual address space of a process.
This enables us to place \verb|f| at a specific virtual address and also separate it from other functions and variables.
On the other hand, in a real attack scenario, the attacker must rewrite an entire page that may contain not only \verb|f| but also other data so that the victim's process does not simply crash.

Figure \ref{fig:timeline_binary} shows the timeline of this experiment.
Here, \verb|A| and \verb|V| refer to the physical addresses that are translated from the virtual address \verb|0x20000| in the address spaces of the attacker and the victim, respectively.
The first and second rows (labeled as {\it attacker} and {\it victim}, respectively) show the operations executed by the two processes.
The third and fourth rows show the address translation information that the two processes would use at a particular point in time.
The victim's process can properly translate \verb|0x20000| to \verb|V| before the attacker conducts GbHammer.
However, after \verb|t = 1s|, the victim's process mistakenly translates \verb|0x20000| into \verb|A|, where the malicious binary blob has been written by the attacker.
This is because the global iTLB entry created by the attacker's process is used by the victim's process as well.

\subsection{Setup: Data Snooping}
In this experiment, the victim's process first allocates a memory region starting from address \verb|0x20000|, 
sleeps for 5 seconds, and then writes a secret string ``This is victim's data'' to the region.
The sleep duration (5 s) can be any value as long as there is enough time for the attacker to do its job between the memory allocation and data writing by the victim's process.
The attacker's process, meanwhile, creates a shared page whose virtual address starts from \verb|0x20000| by GbHammer, reads the data from it every second, and outputs the data concatenated with a string ``This is attacker''.
A {\bf global} dTLB entry is created when the attacker's process reads the shared page for the first time because of our modification to gem5.
The victim's process mistakenly translates the virtual address \verb|0x20000| using this dTLB entry and the secret string is maliciously stored on the shared page.

\subsection{Results}
\begin{figure}[t]
\begin{mylisting}
This is victim.
Expected output: 1
Actual output: 2
\end{mylisting}
\caption{\label{result:binary_result} Result of Binary Execution Experiment}

\vspace{5mm}

\begin{mylisting}
This is attacker
This is attacker
This is attacker
This is attacker
This is attacker
This is attacker
This is attacker This is victim's data
This is attacker This is victim's data
This is attacker This is victim's data
This is attacker 
\end{mylisting}
\caption{\label{result:data_result} Result of Data Snooping Experiment}
\end{figure}

\subsubsection{Binary Execution}
Fig.~\ref{result:binary_result} shows the output from the victim's process we observed on the terminal.
The first and second lines are pre-defined strings and they are properly output.
The value after ``Actual output:'' in the third line is the return value of what the victim's process thinks is \verb|f|.

We confirmed that GbHammer allowed the attacker to make the victim's process execute the binary that the attacker prepared.
After the victim's process executed what it thinks is function \verb|f|, it printed {\bf 2} instead of 1 as shown in Fig.~\ref{result:binary_result}.
This means the binary that the attacker had written to the maliciously shared page was executed instead of \verb|f|.

\subsubsection{Data Snooping}
Fig.~\ref{result:data_result} shows the output from the attacker's process we observed on the terminal.
From the first to the fifth lines (from \verb|t = 0s| to \verb|t = 5s|), the attacker's process printed ``This is attacker'' and nothing after it.
From the seventh to ninth lines (from \verb|t = 6s| to \verb|t = 8|s),
the secret string that the victim had written to a memory region that it had allocated was printed by the attacker.

We confirmed that the GbHammer allowed the attacker to snoop on the victim's data.
After the victim's process had written the secret string to the memory region that it had allocated,
the attacker's process successfully printed that data to the terminal.
This means that the victim was forced to write the secret data to the maliciously shared page and the attacker could read the data from it.

\section{Possible Mitigation Measures}
\subsection{Modifying mmap}
The OS can ignore the virtual address given as an argument of \verb|mmap| to make Step~(2) of GbHammer harder to achieve.
Because the virtual address of an allocated memory region does not matter for ordinary programs,
this change should not affect many use-cases of \verb|mmap|.
For example, when a program maps a file into memory using \verb|mmap|, the same source code of the program should work for any virtual address used as the starting address of the mapping.

There are two use cases of \verb|mmap| we are aware of that require a newly allocated memory region to be placed at a specific virtual address.
Except for these cases (and others that we are not aware of if any), it is safe to simply ignore the virtual address specified as the starting address of the mapping.
The two use cases are as follows:

\begin{enumerate}
    \item {\bf Shared library loading}: Linux programs specify virtual addresses to \verb|mmap| when they load shared libraries (i.e., \verb|.so| files).
    This is because a shared library contains different sections that must be loaded with different protection modes (e.g., read-only, executable-but-not-writable) but must also be contiguous to each other.
    One way of achieving this without specifying the starting addresses of mappings would be to create a new system call that accepts multiple protection modes and offsets inside a mapping to use each protection mode.
    For example, a program can use this new system call to map \verb|lib.so| to any address, and make it read-only from offset \verb|0| to offset \verb|1024| but executable-but-not-writable from offset \verb|1024|.
    \item {\bf Container live migration}: migrating a container from one host machine to another requires memory allocation in the destination host with the virtual addresses specified.
    This is because the layout of the virtual address spaces of the processes that the container consists of must be replicated from the source to the destination host.
    Otherwise, every value in the container's memory and registers that represents an address must be rewritten to a new value.
    This is impractical because there is no decisive method to know if a value (e.g., \verb|0x20000|) is an address or something else.
\end{enumerate}

\subsection{Modifying the Loader}
Specifying virtual addresses of sections in an ELF binary when it is loaded is another method to achieve Step~(2) of GbHammer.
The OS can ignore this as long as the ELF binary is compiled as position-independent code (PIC).
PIC enables sections of an ELF binary to be loaded on any virtual address by placing fake addresses in the binary that are rewritten to actual ones at load time.

This measure is applicable as long as the source code of the victim's program is available and compatible with a modern compiler.
Because creating a PIC binary requires compiling it from the source code, it cannot be applied to existing non-PIC programs whose source code are either closed or lost.

\subsection{Ignoring the Global Bits}
Enabling the global bit of a PTE does not guarantee that the address translation information is used by other processes, as the Intel manual says {\it a processor may use}~\cite{IntelManual} global TLB entries.
The PGE bit (bit 7) in a control register named CR4 decides whether the global bit of a TLB entry is respected or not (described in Section 4.10.2.4 ``Global Page'' of~\cite{IntelManual}).
Thus, the risk of GbHammer is completely eliminated if this flag is disabled.
However, it is important to assess the effectiveness of this measure by considering the overhead incurred by completely disabling the global bits.
As far as we know, this overhead is not well studied under the combination of modern processor architectures and modern applications.

\section{Discussion}
\subsection{GbHammer in ARMv7 and RISC-V}
GbHammer and the exploits based on it are also applicable to the ARMv7 and RISC-V ISAs. A PTE of ARMv7 has nG bit (non Global bit) which functions conversely to the global bit in x86~\cite{ARMManual}. 
Just like on x86, an attacker can share a single page with the victim with each successful GbHammer attempt on an ARMv7 processor.
This is because the nG bit exists only in the last level of the hierarchy of address translation information where the physical page number is stored.

The hierarchy of address translation information in RISC-V is different from that in x86 and ARMv7.
Unlike them, every level of the hierarchy in RISC-V has the same management bits including G bits, meaning that every level can be global.
Specifically, the manual~\cite{RISCVManual} says in its Section~10.3.1 that {\it for non-leaf PTEs, the global setting implies that all mappings in the subsequent levels of the page table
are global}.
Note that every level of address translation information is called a page table in RISC-V (unlike page table, page directory, etc. in x86).
The risk of GbHammer in RISC-V could be much larger than that of x86 or ARMv7, because a much larger address range is maliciously shared at once.

\subsection{Relaxed Constraints on the Victim's Process}
We let the victim's process sleep for some amount of time (5 seconds specifically in our current experiments) so that the attacker's process can execute GbHammer during this period.
This constraint could be a large hurdle for the attacker because there is no straightforward way to let other processes sleep with a user privilege.
Instead, an attacker could target a process that repeatedly accesses the same virtual address (e.g., calling the same function over and over) in a loop to relax this constraint.
This effectively gives the attacker enough time to execute GbHammer to the address that the victim accesses.
In this scenario, the attacker needs to invalidate the non-global TLB entry created by the victim's process by using measures such as constructing eviction sets for the TLB~\cite{Zhang2020}.

Our experimental setup needs improvement to conduct experiments for this scenario.
In the current setup, we set the global bit of a TLB entry when its corresponding virtual address is \verb|0x20000|.
This happens no matter which process creates the TLB entry because our modified gem5 does not distinguish the process that creates a TLB entry.
Thus, a TLB entry is mistakenly set as global when the victim's process first creates one inside a loop,
resulting in a diversion from what would happen when a real GbHammer attack is executed.

\section{Related Work}
GbHammer and the exploits based on it are novel in three aspects.
First, GbHammer reveals the risk of management bits in page table entries being hammered, while previous studies focus on hammering the address parts.
Zhang {\it et al.}~\cite{Zhang2020} show that an attacker can access protected memory regions such as the kernel space by maliciously overwriting address translation information on a page table entry.
Yuan {\it et al.}~\cite{Xiao2016} demonstrate that a bit flip induced by RowHammer attack can bypass memory isolation forced by virtualization.
Second, the arbitrary binary execution exploit is advanced in its uniqueness.
Although many exploits based on RowHammer have been reported~\cite{Zhang2020,Kwong2020,Jang2017,Xiao2016,Li2024}, we are the first to achieve arbitrary code execution as far as we know.
Finally, the data snooping exploit could be advanced in its reliability and read speed.
A previously known exploit that enables reading data is RAMBleed~\cite{Kwong2020}.
Because RAMBleed uses RowHammer to guess the data bits stored in the aggressor row by observing bit-flips incurred to the victim row, it only achieves {\it 0.31 bits/second at an accuracy rate of 82\%}~\cite{Kwong2020}.
Our data snooping exploit can read data with much faster speed and higher accuracy once a shared page is successfully created because a data read from the shared page is merely a normal memory copy (from the shared page to either a register or another memory page).

\section{Conclusion and Future Work}
While various exploits based on RowHammer have been discovered, this research focuses on the risk of the global in a page table entry (PTE) being hammered.
We found that two new exploits, arbitrary binary execution and data snooping, are possible by flipping the global bit with RowHammer.
We also demonstrated that they actually work for a real Linux kernel on a cycle-accurate CPU simulator.
Our future work includes observing the behavior of GbHammer in real machines and reproducing GbHammer in RISC-V to prove that it indeed has a larger risk than in x86 and ARMv7.

\section*{Acknowledgment}
This work was supported by JST Global Science Campus Experts in Information Science,
and JST, PRESTO Grant Number JPMJPR22P1, Japan.
We thank the anonymous reviewers for their valuable feedback to improve this paper.

\bibliographystyle{IEEEtran}
\bibliography{main}

\end{document}